\documentclass[twocolumn,backref, breaklinks, colorlinks]{aastex63}
\hypersetup{linkcolor=blue,citecolor=blue,filecolor=cyan,urlcolor=blue}
\usepackage{tabularx}
\usepackage{url}
\usepackage{graphicx}
\usepackage[T1]{fontenc}
\usepackage{ae,aecompl}
\usepackage{booktabs}
\usepackage{natbib}
\usepackage{enumitem}
\usepackage{mathtools}
\usepackage{graphicx}
\usepackage{amssymb}
\usepackage{amsmath}
\usepackage{ulem}
\usepackage{epstopdf}
\usepackage{color}
\usepackage{graphicx}
\usepackage{xcolor}
\usepackage{needspace}

\usepackage{blindtext}
\usepackage{longtable}
\usepackage{tabu}
\usepackage{lineno}

\def\unit #1{\,{\rm #1}}

\newcommand\kev{\rm \,\unit{keV}}

\newcommand\lunit{\rm \,erg \,s^{-1}}

\newcommand\ledd{L_{\rm Edd}}

\newcommand\lambdaedd{\lambda_{\rm Edd}}

\newcommand\lbol{L_{\rm  bol}}

\newcommand\msol{M_{\odot}}

\newcommand\mbh{M_{\rm BH}}

\newcommand\ks{\, \rm ks}

\newcommand\pc{\unit{pc}}

\newcommand\ev{\unit{\, eV}}
\usepackage{amsmath} % or simply amstext
\newcommand{\angstrom}{\text{\normalfont\AA}}

\newcommand\chandra{{\it Chandra}}

\newcommand\swift{{\it Swift}}
\newcommand\xmm{{\it XMM-Newton}}

\newcommand\suzaku{{\it Suzaku}}
\newcommand\nustar{{\it NuSTAR}}

\def\aap{A\&A} \def\apjl{ApJ}  
 \def\apjs{ApJS}

\accepted{{\bf Review article} in \textbf{Frontiers in Astronomy and Space Sciences}}
\shorttitle{Review on AGN corona}

\begin{document}
\tighten

\title{X-ray properties of coronal emission in radio quiet Active Galactic Nuclei}

\author[0000-0003-2714-0487]{Sibasish Laha}

\affiliation{Astrophysics Science Division, NASA Goddard Space Flight Center, Greenbelt, MD 20771, USA.}
\affiliation{Center for Space Science and Technology, University of Maryland Baltimore County, 1000 Hilltop Circle, Baltimore, MD 21250, USA.}
\affiliation{Center for Research and Exploration in Space Science and Technology, NASA/GSFC, Greenbelt, Maryland 20771, USA}

%\author[0000-0003-2714-0487]{Christopher S. Reynolds}

%\affiliation{Institute of Astronomy, University of Cambridge, Madingley Rise, Cambridge CB3 OHA, UK.}

\author[0000-0000-0000-0000]{Claudio Ricci}
\affiliation{Instituto de Estudios Astrof\'isicos, Facultad de Ingenier\'ia y Ciencias, Universidad Diego Portales, Av. Ej\'ercito Libertador 441, Santiago, Chile}
\affiliation{Kavli Institute for Astronomy and Astrophysics, Peking University, Beijing 100871, China}

\author[0000-0002-6460-0078]{John C. Mather}
\affiliation{Astrophysics Science Division, NASA Goddard Space Flight Center, Greenbelt, MD 20771, USA.}

\author[0000-0003-2714-0487]{Ehud Behar}

\affiliation{Department of Physics, Technion, Haifa, Israel}

\author[0000-0002-6460-0078]{Luigi C. Gallo}
\affiliation{Department of Astronomy and Physics, Saint Mary's University, 923 Robie Street, Halifax, NS B3H 3C3, Canada}

%\author[0000-0003-2714-0487]{Abdu Zhoghbi}
%\affiliation{Astrophysics Science Division, NASA Goddard Space Flight Center, Greenbelt, MD 20771, USA.}
%\affiliation{Center for Space Science and Technology, University of Maryland Baltimore County, 1000 Hilltop Circle, Baltimore, MD 21250, USA.}
%\affiliation{Center for Research and Exploration in Space Science and Technology, NASA/GSFC, Greenbelt, Maryland 20771, USA}

\author[0000-0003-2714-0487]{Frederic Marin}
\affiliation{Université de Strasbourg, CNRS, Observatoire Astronomique de Strasbourg, UMR 7550, 67000 Strasbourg, France}

\author[0000-0000-0000-0000]{Rostom Mbarek}
\affiliation{Department of Astronomy, University of Maryland, College Park, MD, USA} 
\affiliation{Astrophysics Science Division, NASA Goddard Space Flight Center, Greenbelt, MD 20771, USA.}

\author[0000-0000-0000-0000]{Amelia Hankla}
\affiliation{Department of Astronomy, University of Maryland, College Park, MD, USA} 

\correspondingauthor{Sibasish Laha}
\email{sibasish.laha@nasa.gov,sib.laha@gmail.com}

%\date{\today}
%\begin{document}
%\pagerange{\pageref{firstpage}--\pageref{lastpage}} \pubyear{2020}
%\maketitle
%\label{firstpage}

\begin{abstract}
	
 Active galactic nuclei (AGN) are powerful sources of panchromatic radiation. All AGN 
	emit in X-rays, contributing around $5-10\%$ of the AGN bolometric luminosity. The X-ray emitting region, popularly known as the corona, is geometrically and radiatively compact with a size typically $\lesssim 10 \, R_{\rm G}$ (gravitational radii). The rapid and extreme variability in X-rays also suggest that the corona must be a dynamic structure.
  Decades of X-ray studies have shed much light on the topic, but the nature and origin of AGN corona are still not clearly understood. This is mostly due to the complexities involved in several physical processes at play in the high-gravity, high-density and high-temperature region in the vicinity of the supermassive black hole (SMBH). It is still not clear how exactly the corona is energetically and physically sustained near a SMBH. The ubiquity of coronal emission in AGN points to their fundamental role in black hole accretion processes. In this review we discuss the X-ray observational properties of corona in radio quiet AGN.

    %The ubiquity of coronal emission in AGN may suggest that it is a fundamental component of black hole accretion process.
%However, the growing consensus from reverberation and polarimetric studies favor an equatorially extended structure or it may even be a part of the inner accretion disk.

\end{abstract}


\vspace{0.5cm}

%\linenumbers

%%%%%%%%%%%%%%%%%%%.  Sec 1.   %%%%%%%%%%%%%%%%%%%

\section{Introduction}\label{sec:intro}

Some of the most energetic emission in an active galactic nucleus (AGN) hosting an accreting supermassive black hole (SMBH) is produced in the X-rays. The AGN corona which is responsible for most of the X-ray emission, is 
an extremely hot ($T\sim 10^9$K) plasma residing very close to the SMBH. The coronal X-ray spectrum is a power-law in the energy range $\sim 0.3-100\kev$ \citep{vaiana1978,haardt1993,merloni2003}, and contributes to around $5-10\%$ of AGN bolometric luminosity \citep{elvis1994, marconi2004,vasudevan2007,fabian2017}. Over the past 40-50 years of X-ray observations, important discoveries have been made in AGN coronal physics, which have opened up new fundamental questions, such as: 1) What is the structure and extent of the corona, and how is it sustained in the high gravity regime? 2) What determines the fraction of thermal and non-thermal electron components in the corona? 3) How is energy pumped and dissipated in the corona? Is the corona in radiative equilibrium?

 %%%%%%%%%%%%%%%%%%%%%%%%%%. Fig 1
 \begin{figure*}
  \centering 

\includegraphics[width=18.1cm,height=8.1cm,angle=0]{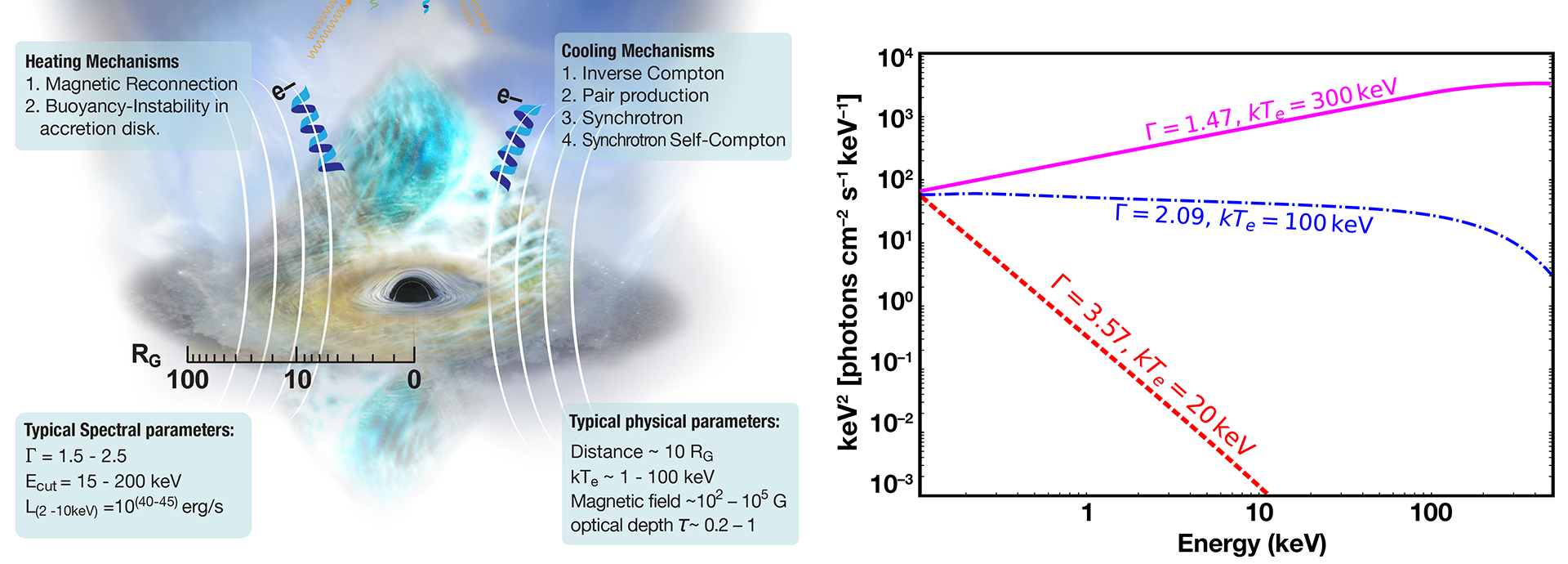}

\caption{\textbf{AGN coronal typical parameters and spectra.} \textbf{ Left Panel:} A cartoon of an AGN central engine with the SMBH (black), accretion disk (brown), the X-ray emitting corona (diffuse cyan), the larger scale outflows (blue) and typical electron orbits around the magnetic fields (helical strips). $R_{\rm G}$ refers to the gravitational radius of the SMBH. The magnetic field lines are shown in white. {\it Top left and right:} dominant heating and cooling mechanisms \citep[see for e.g.,][and references therein]{done1989,balbus1998,dimatteo1998,miller_corona_2000,fabian2015}. {{\it Lower left:} the typical ranges in the spectral parameters: X-ray power law slope ($\Gamma$), power law cut-off energy ($E_{\rm cut}$) and $2-10\kev$ luminosity ($L_{2-10\kev}$). {\it Lower right:} typical ranges in the coronal physical parameters: Distance from the SMBH, electron temperature ($kT_{\rm e}$), Magnetic field in Gauss, and X-ray optical depth ($\tau$). See for e.g., \cite{ricci2017,ricci2018,kamraj2022}.  The typical spectral and physical parameter ranges quoted here represent the bulk of the Radio-Quiet AGN population, though values beyond these ranges are also detected in some AGN.} We also note that the lower limit on AGN X-ray luminosity quoted here corresponds to the detection threshold of current generation instruments and not an intrinsic AGN limit.  \textbf{Right panel:} The shape of the power law continuum for different coronal electron temperatures \citep[for a full discussion see][]{ricci2018}. We used the COMPPS model in XSPEC \citep{arnaud_xspec1996} to simulate Comptonization spectra in X-ray corona. We assumed the following: (a) A spherical geometry of the corona, (b) the plasma optical depth $\tau=0.8$, (c) the disk (UV) seed photons having a  temperature of $10\ev$. We normalized the three spectra at $0.1\kev$. The resulting slope of the power law spectra are reported. We note that with increasing electron temperature the spectral slope becomes harder.}\label{fig:corona_cartoon}

\end{figure*}

 %%%%%%%%%%%%%%%%%%%%%%%%%%%%%%%%%%%%%%%%%%%

The central engine of AGN (see Figure \ref{fig:corona_cartoon} left panel) is thought to consist of an accretion disk surrounding the SMBH. The loss of gravitational energy of the accreting material is expected to be one of the main sources of the energy in AGN, part of which is manifested in the optical and UV bands \citep{shakura1973}. The rate at which the system is accreting is often parametrized as the Eddington ratio ($\lambdaedd$) \footnote{ $\lambdaedd = \lbol/\ledd$, where $\lbol$ is the bolometric luminosity and $\ledd$ is the Eddington luminosity.}.  

In this review we will focus on the coronal X-ray emission from radio quiet AGN (RQ-AGN), which represent the largest population of accreting SMBHs. We do not discuss radio loud AGN (RL-AGN) in this review because the jets may contribute to the X-rays adding extra complexities and contaminate X-ray emission from the corona. We note that in a short review of a mature field such as AGN coronal emission, it is not possible to cover all topics related to the subject, and some subjectivity may unintentionally introduce bias.

This manuscript is arranged as follows: In Section 1.1 we discuss some of the most important physical processes in AGN coronae. In Section 2 we list the phenomenology of the coronal emission discussing the most relevant discoveries and the empirical relations between X-ray coronal emission and the other observable quantities. In Section 3  we briefly address some of the open questions in the field, and in Section 4 we discuss future perspectives.

%
%%%%%%%%%%%%%%%%%%%%%%%%%%%%%%%%% Fig 2

\begin{figure*}
  \centering 

\hbox{
\includegraphics[width=8.5cm,angle=0]{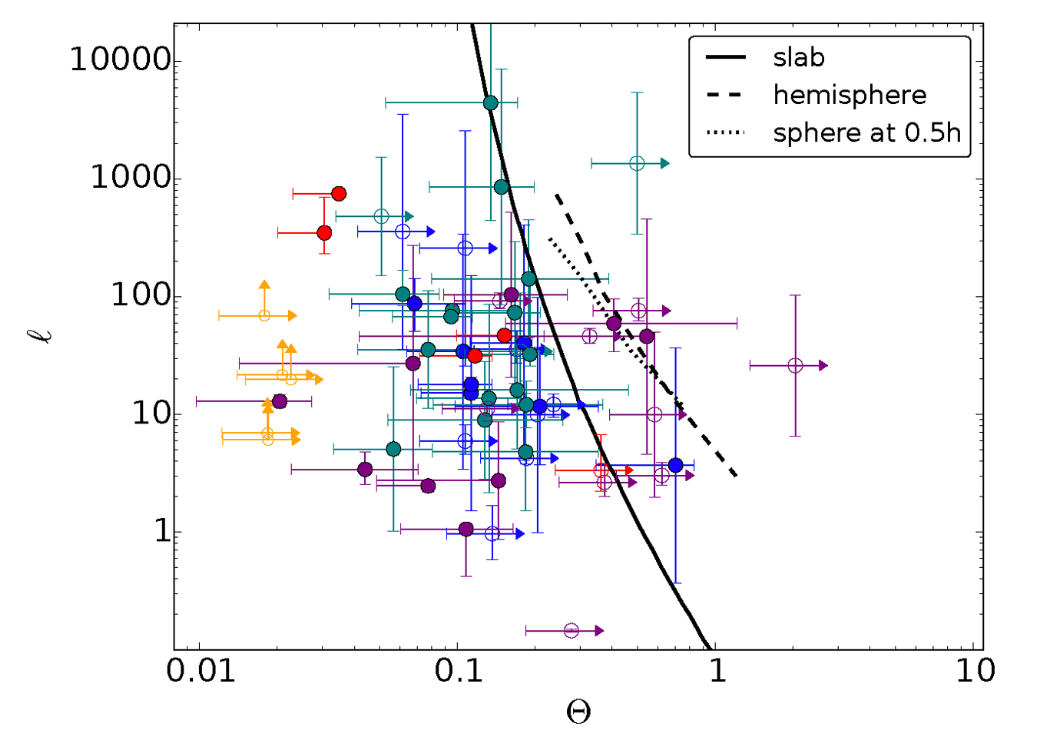}
\includegraphics[width=8.5cm,angle=0]{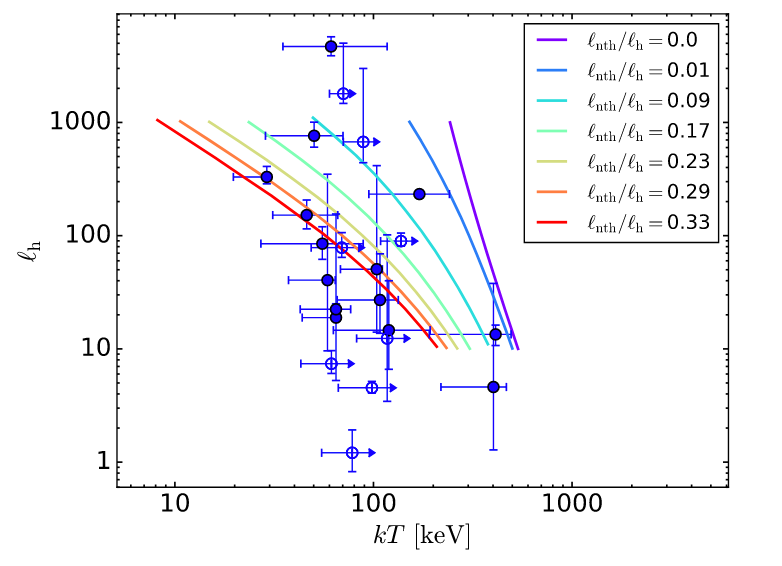}
}
\caption{\textbf{Pair production acting as a thermostat for coronal plasmas.} {\it Left:} Compactness parameter ($l$) and temperature ($\Theta$) phase space plotted for AGN that have measured cut-off energies in the X-ray band \citep{fabian2015,fabian2017}. The black, dashed and dotted lines are theoretical curves denoting the limits for the pair production process for a slab, hemisphere and spherical geometries, respectively. {\it Right:} Same as left, but now the different colored curves denote varying contributions from thermal and non-thermal plasmas, with the non-thermal fraction increasing from right (purple curve) to left (red curve).   
} \label{fig:fabian_pair_production}

\end{figure*}

%%%%%%%%%%%%%%%%%%%%.  Sec 1.1

\subsection{The primary physical processes in AGN corona}

The coronal X-ray emission can be simply characterized by a power law with a photon index ($\Gamma$) and cut-off energy ($E_{\rm C}$), such that the flux $F(E)$ $\propto E^{-\Gamma}e^{-E/E_{\rm C}}$. 
Some of the main observables of X-ray coronae that can be inferred from X-ray spectroscopy are: (1) the spectral slope $\Gamma$, which is related to the temperature of the Comptonizing electrons ($kT_{\rm e}$) and the optical depth ($\tau$) of the X-ray emitting plasma \citep[][]{rybicki1979}, (2) the high energy cut-off $E_{\rm C}$ and (3) the X-ray luminosity $L_{0.3-100\kev}$.
We briefly discuss some of the most important physical processes that are expected to take place in AGN coronae, and we refer the readers to \cite{rybicki1979} for a detailed exposition.

%%%%%%%%%%%%%%%%%%%%%. Inverse Compton

\textbf{Inverse Compton (IC) scattering} is thought to be the dominant process responsible for X-ray emission in AGN. When UV seed photons from the accretion disk, with energies $h\nu$, pass through the coronal plasma, energy gets transferred from the hot ($10^9$ K) electrons to the photon field by repeated IC scattering. This happens if $h\nu <4kT_{\rm e}$, where $h$ is the Planck constant, $\nu$ is the photon frequency, $k$ is the Boltzmann constant and $T_{\rm e}$ is the temperature of the electrons. 

 For a plasma of non-relativistic electrons in thermal equilibrium with energy $kT_{\rm e}<<m_{\rm e}c^2$, having an optical depth $\tau_{\rm es}$, one can define the Compton y parameter: $y= {\rm max}(\tau_{\rm es},\tau_{\rm es}^2)(4kT_{\rm e}/m_{\rm e}c^2)$, where $m_e$ is the electron mass and $c$ is the speed of light. For $y>>1$, the average photon energy reaches the thermal energy of the electrons and is called `saturated inverse Compton scattering'. The case for unsaturated Comptonization ($y >1$) is however of most interest in black hole systems, and in such a case, 
the output spectrum in the X-ray is a power law with a high energy cut-off $E_{\rm C}$ determined by the electron temperature, which is typically approximated to be $E_{\rm C}\sim 2-3kT_{\rm e}$ \citep{petrucci2001,fabian2015}.  %\textbf{\color{red} [1] Chris Reynolds comments: Need to introduce this process a little more... See Reynolds and Nowak 2003}. \\

%%%%%%%%%%%%%%%%%%%%%. Synchrotron

 \textbf{Synchrotron emission:} The high densities of electrons around the magnetic field in AGN corona makes it a significant synchrotron emitter predominantly between $5-200\, \rm GHz$ \citep{laor2008,panessa2019,baldi2022,kawamuro2022,ricci_radio_correl_2023}. The fact that (1) we see ubiquitous unresolved mm emission even with high spatial resolution, (2) flat radio slopes, and (3) strong correlation between the radio and X-rays in RQ-AGN are indicators of radio emission from the corona through synchrotron processes \citep{panessa2019}. For example, recent results \citep{ricci_radio_correl_2023} point toward a tight correlation between $2-10\kev$ and $100$ GHz luminosity for a volume-limited sample of nearby hard X-ray selected RQ-AGN. Similarly, the core radio flux at 5 GHz and the $2-10\kev$ luminosity for nearby radio quiet AGN have been found to show an interesting correlation $L_{\rm R,\, 5GHz}$/$L_{2-10\kev}\sim 10^{-5.5}$ \citep{laor2008} which is similar to that found in coronally active stars (such as the Sun) and is popularly known as the Gudel-Benz relation \citep{gudel_benz1993}. As a caveat we note here that the coronal magnetic field can be as high as $B=10^2- 10^5$ Gauss and significant synchrotron self absorption (SSA) effects may limit our detection at lower frequencies (below $\sim 40$ GHz). 
 
 Direct measurement of magnetic fields in AGN corona has not yet been possible, but we can estimate a typical range from the analogy of RL-AGN. Large $B_0$ values on event horizon scales are feasible considering that measurements from AGN jets have previously found magnetic field strengths of $\sim0.1$ G on $\sim1$ pc scales from core frequency-shift methods~\citep{osullivan2009} and $\sim10$ G on $\sim0.1$ pc scales from Faraday rotation measurements~\citep{marti-vidal2015}.
Such observational values are consistent with $B_0 \gtrsim 10^5$~G at the base of the jet with a $1/r$ decay of the magnetic field, and are thereby consistent with theoretical and numerical predictions for launching relativistic jets~\citep{tchekhovskoy2011}.

 %%%%%%%%%%%%%%%%%%% Pair production
 
\textbf{Electron-positron pair production:} Sources which are physically compact and highly luminous, like the AGN corona, are also radiatively compact. This means that the photons and the particles in the plasma are in constant interaction with each other. In such a plasma, photon–photon collisions can lead to $e^{-}-e^{+}$ pair production, when the photons are energetic enough. The resulting $e^{-}-e^{+}$ pair density is proportional to the luminosity ($L$) and  electron temperature ($kT_{e}$ or $\Theta = kT_{e}/ m_{e}c^2$), and inversely proportional to the source size ($R$) assuming a spherical source. This is typically expressed by the  compactness parameter $l=L\sigma_{\rm T}/(R m_e c^3)$ where $\sigma_{\rm T}$ is the Thomson cross-section. Thus, when the energy content of corona increases, manifested by an increase in both $l$ and $kT_{\rm e}$, the extra energy goes into creating more pairs, rather than increasing the temperature. Therefore, the process acts as a natural thermostat for the corona\citep{done1989,fabian2015,fabian2017}. See Figure \ref{fig:fabian_pair_production} left panel. 

If magnetic reconnection is a dominant form of energy production mechanism in the X-ray corona \citep{dimatteo1998}, then one would expect a fraction of non-thermal electrons \citep{done1989}. The existence of non-thermal particles in the corona would result in a distribution of photon energy that extends into the the MeV band. This small number of high energy particles could be highly effective in seeding pair production. Moreover, the cooled non-thermal pairs could share the total available energy, thus reducing the mean energy per particle and therefore decreasing the temperature of the thermal population. Such hybrid coronal plasma, consisting of thermal and non-thermal electron populations, might have been found in a few nearby AGN (See Figure \ref{fig:fabian_pair_production} right panel), in which the Comptonizing plasma is found well below the  pair production line in the $l-kT_{\rm e}$ plane \citep{fabian2015,fabian2017}.

%%%%%%%%%%%%%%%%%%%.  Sec 2.   %%%%%%%%%%%%%%%%%%%

\section{Phenomenological properties of the corona}

 X-ray emission from AGN was detected and studied already by the early X-ray observatories such as {\it Ariel-V} (1974-1980 \citealt{smith1976}), {\it HEAO-1} (1977-1983, \citealt{rothschild1979}), {\it HEAO-2} or {\it Einstein} (1978-1981, \citealt{giacconi1979}), {\it EXOSAT} (1983-1986, \citealt{taylor1981}), {\it GINGA} (1987-1991, \citealt{makino1987}). In the later period {\it ASCA} (1993-2001, \citealt{tanaka1994}) and {\it RXTE} (1995-2012, \citealt{Swank1999}) provided seminal insights into the X-ray properties of the AGN corona. For example, the ubiquity of X-ray emission from Seyfert-1 galaxies was established \citep{elvis1978} by the first catalog from the Ariel-V sky survey \citep{cooke1978}. The first large spectral samples of AGN observed by HEAO-1 revealed that the observed range in photon spectral indices was tightly distributed around $\Gamma\approx1.7$ \citep{mushotzky1980,rothschild1983,mushotzky1984}. The {\it Einstein} and {\it EXOSAT} missions demonstrated that rapid, large amplitude X-ray variability is a common feature in nearby AGN, and that such variability is stochastic and it shows no characteristic timescale \citep{lawrence1987,mchardy1987}. 

These discoveries were followed by the era of the great X-ray observatories, which started with the launch of  \chandra{} (1999-, \citealt{Weisskopf1996}) and \xmm{} (1999-, \citealt{Lumb2012}), and later with the advent of hard X-ray ($>10\kev$) observatories such as {\it INTEGRAL} (2002-, \citealt{winkler2003}), \swift{}-BAT (2004-, \citealt{barthelmy2005}), \suzaku{} (2005-2015, \citealt{mitsuda2007}) and \nustar{} (2012-, \citealt{Harrisson2013}). Our understanding of AGN corona over the years has improved significantly, but much remains to be understood. In this section, we will review some of the most important observational characteristics of AGN coronal emission obtained with the above mentioned observatories.

%%%%%%%%%%%%%%%%%%%.  Sec 2.1  %%%%%%%%%%%%%%%%%%%

\subsection{The coronal plasma spectral and physical properties.}

\begin{figure*}
  \centering 

\hbox{
\includegraphics[width=10cm,angle=0]{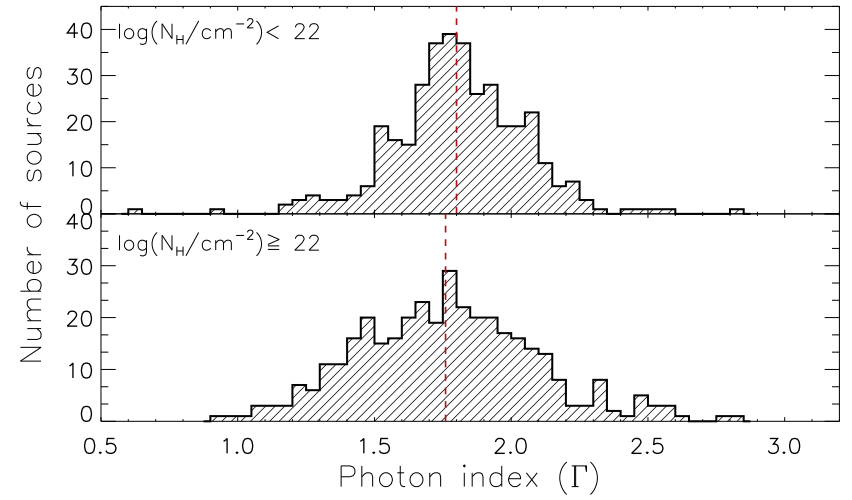}
\includegraphics[width=8.5cm,angle=0]{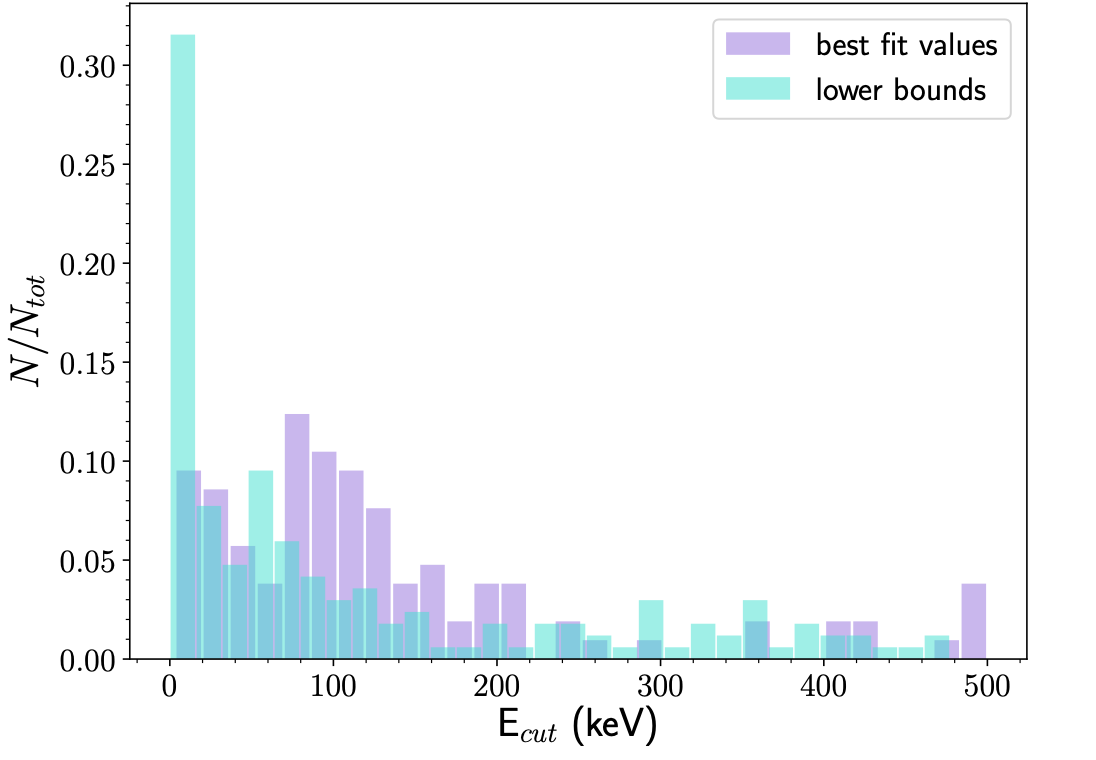}

}\caption{\textbf{Distribution of the X-ray spectral parameters of AGN coronal emission:} {\it Left:} The distribution of the photon-indices $\Gamma$ for X-ray unobscured (top panel) and X-ray obscured (lower panel) sources \citep{ricci2017}.  {\it Right:} Distribution of the cut-off energy as measured for a sample of AGN with \nustar{} hard X-ray observations \citep{kamraj2022}, when the spectra are modeled using a simple absorbed power law. The cyan and purple histograms represent the lower bounds and the best-fit values, respectively.} \label{fig:distributions}

\end{figure*}

\textbf{Coronal X-ray power law slope and optical depth:} Figure \ref{fig:corona_cartoon} left panel highlights the primary spectral and physical characteristics of X-ray coronae and their typical range of values. The left panel of Figure~\ref{fig:distributions} shows the photon index ($\Gamma$) distribution for a large sample of AGN (both obscured and unobscured) studied with broad-band X-ray observations (0.3--150\,keV), with a median value of $1.78\pm0.1$ \citep{ricci2017}. Recent broad-band X-ray studies also show that the spectral slope of the $14-195\kev$ emission is steeper than the $0.3-10\kev$ band, suggesting the high energy cut-off is ubiquitous in AGNs \citep{ricci2017}.

The photon index is dependent on both the plasma temperature and the optical depth, and it can be estimated as 
$\Gamma\sim[\frac{9}{4} + \frac{m_ec^2}{kT_e\tau (1+\tau/3)}]^{0.5} -\frac{1}{2}$ \citep{rybicki1979}. By measuring both $\Gamma$ and $E_{\rm C}$ it is possible to estimate the optical depth of the Comptonizing plasma, assuming a geometry \citep[see for example][]{brenneman2014}. Recent studies of nearby AGN estimate a median value of the optical depth of $\tau=0.25\pm 0.06$ \citep{ricci2017}.

%%%%%%%%%%%%%%%%%%%%%%%%%%%%%%%%%%

\textbf{Coronal Luminosity and bolometric correction:}  The typical coronal luminosity can span a large range $L_{2-10\kev}\sim 10^{40-45}\lunit$ \citep{piconcelli2005,She17,ricci2017}, with the lower limit being only loosely defined by detector sensitivity and increasing contribution of non-AGN process to the X-ray emission. On this low-luminosity end of the distribution are the sources that are either accreting at low Eddington ratios or host intermediate mass black holes ($\log(\mbh/\msol) < 6$, \citealp{dong2012}). On the extreme high-luminosity end, are the hyperluminous AGN with  $L_{2-10\kev} \ge 10^{45}\lunit$, typically found at $z\sim 2-4$ encompassing the cosmic peak of quasar activity \citep{martocchia2017}. 

The contribution of the X-rays to the total AGN emission is usually parametrized with the X-ray bolometric correction ($\kappa_{\rm 2-10}):$ $\kappa_{\rm 2-10} \sim\lbol/L_{\rm 2-10}$.  
Studies of nearby AGN have shown that more luminous sources typically have weaker coronal X-ray emission relative to their bolometric luminosity, with $\kappa_{\rm 2-10}\simeq 15-25$ at $\lambdaedd<0.1$, and $\kappa\sim 40-70$ at $\lambdaedd>0.1$  \citep{vasudevan2007}. 

%%%%%%%%%%%%%%%%%%%%%%%%%%%%%%%%%%%%%%%%%%%%%

\textbf{The high energy cut-off ($E_{\rm C}$) and the coronal temperature:}  { The high energy cut-off of the power law component is related to the coronal temperature as $E_{\rm C}\sim 2kT_{e}$, for an optically thin plasma, i.e., $\tau \lesssim 1$. On the other hand when the plasma is optically thick, i.e., $\tau>>1$ the relation is $E_{\rm C}\sim 3kT_{e}$, both approximated for a corona of slab geometry \citep[][]{petrucchi2000,petrucci2001}.} The right panel of Figure \ref{fig:distributions} \citep{kamraj2022} shows the distribution of high-energy cut-offs inferred from \nustar{} observations of a sample of nearby AGN. The median value of the cut off energy obtained for the sample is $\sim 84\pm 9 \kev$. A large study of a sample of nearby \swift{}-BAT detected AGN  finds a median cut-off energy in local AGN that is significantly higher ($E_{\rm C}\sim 200\pm29\kev$; \citealp{ricci2017}). Indirect constraints on the cut-off energy have been obtained by fitting the cosmic X-ray background (CXB), and have shown that the mean cut-off energy is likely below $300\kev$\citep{gilli2007,treister2009,ueda2014}, in agreement with the observational studies reported above. 

Analyzing a sample of $\sim 200$ AGN, \citet{ricci2018} found that, while $E_{\rm cut}$ is not related to the mass of the black hole or the $14-150\kev$ luminosity, it appears to be related to the Eddington ratio ($\lambdaedd$). Sources with $\lambdaedd>0.1$ were shown to display significantly lower median cut-off energy ($E_{\rm cut}=160\pm 41\kev$) than those with $\lambdaedd\le 0.1$ ($E_{\rm cut}=370\pm 51\kev$). This supports the idea that more radiatively compact coronae are cooler, because they tend to avoid the region in the temperature-compactness parameter space where runaway pair production would dominate (See Fig \ref{fig:fabian_pair_production}).  
 In some extreme cases, coronal temperatures as low as $kT\sim 10-20\kev$ have been measured in a few nearby AGN \citep{buisson2018}. Interestingly, and in agreement with the results of \citet{ricci2018}, such cool corona are often detected in several high Eddington AGN \citep{kara2017,tortosa2022}. These plasma are either not pair-production dominated, or they are hybrid, as discussed in the Introduction. 
 
  Sample studies of AGN in hard X-rays  with \nustar{} detected an anti-correlation between $kT_{\rm e}$ and $\tau$ \citep{tortosa2018,kamraj2022,serafinelli_corona_2024}. On average, the lower mass, highly accreting  narrow line Seyfert 1 galaxies (NLSy1s) exhibit a steeper photon index ($\Gamma>2$), suggesting the corona might be cooler or less optically thick compared to other AGN (e.g. \citealt{brandt1997,gallo2018}).

%%%%%%%%%%%%%%%%%%%.  Sec 2.2   %%%%%%%%%%%%%%%%%%%

\subsection{The coronal size, geometry and stability.}

 Although corona is known to be compact \citep{ghisellini2004,fabian2015}, it can sometimes be patchy (e.g. \citealt{haardt1991,Stern95,petrucci2013, WG15comp}). Four (simplified) coronal geometries that are commonly discussed in the literature: a point source, a cylindrical slab, a spheroid/ellipsoid, and a conical geometry \citep{gonzalez2017}. Ray-tracing simulations suggest that some of these geometries could be distinguished through X-ray spectral modelling (e.g. \citealt{Wilkins2012b, dauser13, gonzalez2017}) and polarization studies (e.g. \citealt{schnittman2010_xray_polarisation, Zhang19}).  

Although the geometry of the corona is extremely hard to determine, the size of the corona can be inferred from several indirect methods:

%%%%%%%%%%%%%%%%%%%%%%%%%%%%%. 2. X-ray reflection features

\textbf{(1) Spectral and spectral-timing techniques:}\\

\textbf{Emissivity profile:} The emissivity profile describes the amount of reprocessed radiation emitted from the disc as a function of distance from the illuminating source, and it is typically inferred from the properties of the relativistically broadened emission lines (e.g., Fe~K$\alpha$).  The emissivity profile is dependent on the morphology of the corona and its height above the disc (e.g. \citealt{wilkins2011, Wilkins2012b, dauser13, gonzalez2017}). Measurements of the emissivity profile in a few well studied AGN suggest that the corona is relatively compact ($\lesssim 10 \rm R_G$, \citealt{wilkins2014,wilkins2015}).

 \textbf{Reflection fraction:} The detection of broad (and redshifted) FeK$\alpha$ emission lines and its variability in some sources clearly indicates the presence of general relativistic effects in producing the line shape, which may arise out of reflection from the inner regions of an accretion disc \citep{miniutti2004}. The ratio of the reflected flux (primary flux reflecting off the disk) and the primary X-ray flux can provide some constraints on the location and motion of the corona \citep{wilkins_gallo2015,dauser2016,gonzalez2017}. For example, a detailed spectral analysis of the low-flux state of Mrk 335 by \nustar{} revealed a spectra with a high reflection fraction ($>8$) indicating relativistically blurred emission, from a X-ray point source (corona) collapsing down to within $\sim 2 R_{\rm G}$ of the SMBH event horizon. Later on with increasing X-ray flux, the reflection fraction decreased, consistent with a corona moving up to $10\, R_{\rm G}$ as the source brightened \citep{parker2014}.

%%%%%%%%%%%%%%%%%%%%%%%%%%%% Fig 4
%%%%%%%%%%%%%%%%%%%%%%%%%%%%%%%%%%% Fig 3

\begin{figure*}
  \centering 

%\hbox{
\includegraphics[width=18cm, height=8cm]{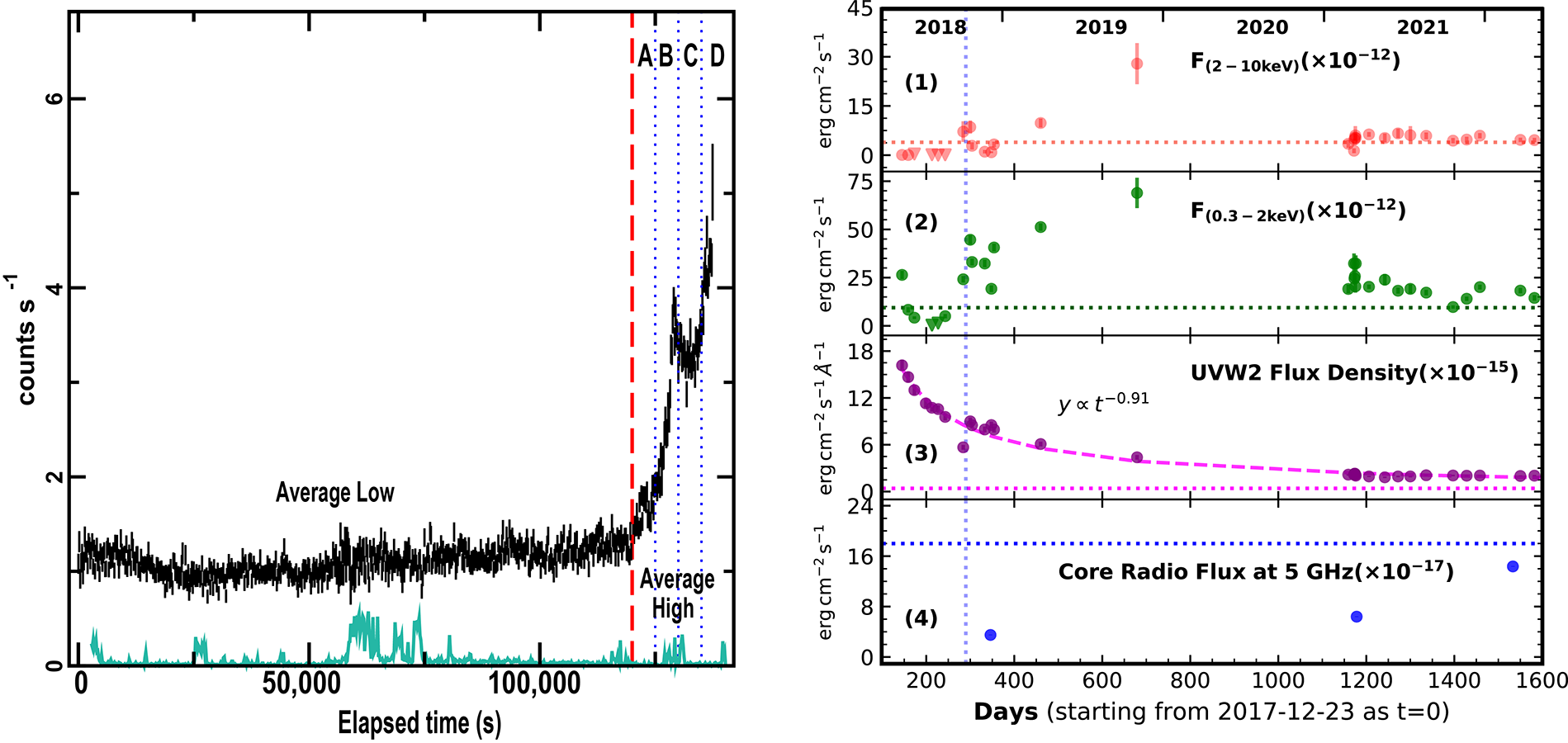}

%\includegraphics[height=8.5cm,width=9.5cm,angle=0]{lc02 copy.pdf}
%\includegraphics[height=9.5cm,width=9.5cm,angle=0]{1es_figure1_main_review.pdf}
%}
\caption{ \textbf{The short and long term X-ray flares in AGN corona.} {\it Left:} The short term ($\lesssim 100 \ks$) X-ray flare in the type-1 AGN Mrk~335 captured by \xmm{} during the rise \citep{gallo2019}. The flux increased by a factor of $\sim 5$ in $20\ks$. The $2-10\kev$ coronal spectral slope $\Gamma$ show gradual softening during the flare-rise (soft-when-bright behavior), estimated for the time bins A, B, C and D \citep{gallo2019}. The black curve is the source+background light curve, while the cyan curve is the background light curve. {\it Right:} The long term X-ray, UV and radio light curve of the changing-look AGN 1ES~1927+654 spanning four years (2018-2022). From top to bottom are: (1) X-ray $2-10\kev$ flux, (2) X-ray $0.3-2\kev$ flux, (3) The UVW2 flux, (4) The core radio ($<1\pc$) flux at 5GHz. The top three panels are from \swift{} observatory, while the radio data are from VLBA. While both the soft and the hard X-rays first vanished (vertical dotted line) and then flared by a factor of $\sim 10$ in $1$ year, the UV showed a constant decline with a powerlaw fall ($\propto t^{-0.91}$) indicating no correlation between the X-rays and the UV. The core radio flux was at its lowest when the X-rays were at its lowest state indicating a close connection between these two bands \citep{laha_1ES_2022,ghosh2023}. The vertical dotted line denotes the time when the X-rays revived. The vertical dotted line denotes the time when the X-rays revived. The horizontal lines in every panel refer to the pre-changing-look values obtained in May 2011.} \label{fig:flare}

\end{figure*}

\textbf{X-ray variability:}
Coronal X-ray emission shows variability at different time scales ($\delta t$), from a few 100 seconds to days \citep{mchardy2005,middei2022,reeves2021}. The shortest variability timescales put an upper limit to the size of the emitting region $R<c\delta t \simeq 1-10 \, R_{\rm G}$. Variability in the coronal emission is echoed in the emission reflected in the accretion disc, with some delay that corresponds to the light travel time between the corona and the disk (e.g \citealt{fabian2009,zoghbi2010,uttley2014}). 
These reverberation lags can provide insights on the location of the corona relative to the inner disc. For sources where reverberation lags have been detected, indications are that the region is compact and typically less than $\sim10 R_{\rm G}$ \citep{demarco2013,kara2016, wilkins-fabian2013,cackett2014, kara2013_xray_reverb_lag,zoghbi2012}.

%%%%%%%%%%%%%%%%%%%%%%%%%%%%. 1. Polarization:
\textbf{(2) X-ray polarization:} Compton scattering induces polarization of the X-ray photons, which is an important tool to study the geometry of the emitting plasma. The polarization of the X-ray photons measured both in degree and position angle, is energy- and geometry dependent. For example, a polarization degree of $4\%$ or higher can possibly rule out spherical and lamp-post coronal models, because symmetry reduces the polarization degree \citep{Zhang19,ursini2022}. The same is true for the orientation of the polarization angle, that is model-specific (different models predict different values of polarization angle).

Launched in December 2021, the Imaging X-ray Polarimetry Explorer (IXPE), is the first X-ray spectro-imaging polarimeter satellite sensitive in the 2-8 keV band \citep{Weisskopf2022}. IXPE successfully measured a polarized signature in NGC~4151 \citep{Gianolli2023} with a polarization degree of 4.9\% $\pm$ 1.1\% at a position angle of 86$^\circ$ $\pm$ 7$^\circ$ east of north at 68\% confidence level. The amount of polarization associated with the corona is of the order of 4 to 8\%, which directly excludes a spherical geometry \citep{behesthipour2017,Ursini2020pol}. The polarization angle measured for this source, which is parallel to its radio jet, suggests that the corona could be distributed along the accretion disk or perhaps it's a part of the inner accretion flow (as in a slab geometry).  On the other hand, upper limits (at 99\% confidence level) of 3.2\% and 6.2\% were obtained for the polarization degree in MCG$-$5$-$23$-$16 \citep{Marinucci2022,Tagliacozzo2023} and IC\,4329A \citep{Ingram2023},  respectively, implying that, for those two objects, we cannot directly rule out a spherical and/or lamppost corona. 
{ However, the orientation of the polarization angle in those two AGN seems to be more consistent with an extended corona (along the equatorial plane) rather than with a polar or spherical corona, because their tentatively measured polarization angle (at $<3\sigma$) are also parallel to the detected radio structures and polar winds \citep{Tagliacozzo2023,Ingram2023}.

Here we also mention an interesting result from an X-ray binary, Cygnus X-1, for which the polarization degree could be constrained exceptionally well,  at (4.0 $\pm$ 0.2)\% between 2$-$8~keV, and a polarization angle parallel to the jet axis \citep{Krawczynski2022}. This suggests that, similar to AGNs, the hot X-ray corona is likely spatially extended in a plane perpendicular to the jet axis, parallel to the inner accretion flow, and rules out the commonly used lamp-post model.

%%%%%%%%%%%%%%%%%%%%%%%%%%%%%. 5. Micro-lensing studies

\textbf{ (3) Microlensing studies:} Gravitational microlensing of quasar light by a foreground mass (lens) can be used to probe the sizes related to the accretion disc and corona (e.g. \citealt{chartas2009,morgan2008,dai2010}).  Using such methods, the size of the X-ray emitting region (the corona) is estimated to be very compact, around $5-10\, R_{\rm G}$ (e.g. \citealt{dai2010}).

%%%%%%%%%%%%%%%%%%%%%%%%%%%%%.6. X-ray eclipses:

\textbf{(4) X-ray eclipses:}   Capturing the transit of the X-ray source by an obscuring cloud is fortuitous, but not rare \citep[see for e.g.,][]{risaliti07,Turner18, Gallo21,ricci2022}. Such events are important as they can be used to constrain the sizes of the X-ray region based on the duration of the eclipse. This method assumes that the cloud is gravitationally bound to the central SMBH in a Keplerian orbit and the eclipse occurs when the cloud moves across our line of sight to the central engine.  In the objects for which eclipses could be used to measure the size of the corona (e.g. \citealt{risaliti2011, Gallo21}), the results have been consistent with those obtained using other methods.

\subsection{Coronal X-ray variability and flares}

AGN coronal X-ray emission is variable at different timescales and with different amplitudes \citep[see for e.g.,][and references therein]{mchardy1987,mushotzky1993,papadakis2004,Mchardy2004,mchardy2006,serafinelli_corona_2024}. Here we briefly discuss three types of AGN X-ray coronal variability commonly observed: (a) Stochastic variability, (b) quasi-periodic variability and (c) Flares, and we note that a detailed discussion of timing and spectral-timing studies of AGN corona is beyond the scope of this short review.\\

\noindent\textbf{(a) Stochastic variability:} One of the most common coronal variability pattern is the chaotic total intensity variation, or stochastic variation. Early studies using observations from Ariel-V and EXOSAT show that $40\%$ of AGN exhibit stochastic variability on a timescale less than 1 day, and 97\% of them showed variability on longer timescales (e.g., \citealp{grandi1992, mchardy1987}). More recently, a linear relationship between the rms amplitude of short-term variability and flux variations on longer timescales has been found in AGN X-ray light curves \citep{gaskell2004,uttley2005,vaughan2011}. This has been dubbed the `rms-flux' relation \citep{uttley2011}. This is an important feature of the aperiodic variability of accreting compact objects, including black hole X-ray binaries \citep{gleissner2004,Heil2012}. 

It is still unclear how the X-ray coronal variability at different-timescales is produced. Popular models predict that inward propagation of random accretion rate fluctuations in the accretion flow could create such stochastic variations in the coronal X-ray emission \citep{lyubarskii1997,kotov2001,king2004,kelly2010,Ingram2013,cowperthwaite2014}. The longer term variability may be produced by accretion rate changes \citep{mushotzky1993}, but the origin of the  short timescale variations ( a few $\ks$ or less) are still debated. 
The magnetic field reconnection in accretion disk threading the coronal plasma likely play a role \citep{dimatteo1998} as they do in the solar corona.

 An important measurement of the variability is the  power density spectrum \citep{Vanderkliss1989,vaughan2003a, vaughan2003b}, which describes the amount of power (the amplitude squared, i.e., the power of the signal) as a function of temporal frequency.
 When the X-ray light curve can be described as random displacements around a mean value, then the power density spectrum (PSD) shows a constant value, that is, all frequencies have equal power. This is known as a white noise spectrum. On the other hand, a red noise spectrum is created when the points in the light curve have a random displacement from its adjacent point rather than from the mean. In such a case the variations at  lower frequencies have more power. Red noise is the characteristic of several astrophysical systems including the Sun \citep{Lu_1991_solar_avalanche} and black hole binaries \citep{belloni1990}, and it is closely related to the stochastic nature of such non-linear systems. In AGNs, over the frequency range $f=10^{-3}$ to $10^{-5}$ Hz, the power spectral density of most Seyfert galaxies has a mean slope of $\alpha \sim 2.0$ in the $2-10\kev$ band, exhibiting no characteristic timescales \citep{gonzalez-martin2012}, and indicating that red-noise steeply decreases at higher frequencies (that is shorter time scales). In some AGN there is a break in the PSD slope at  $f=2\times 10^{-4}$\,Hz, from a much flatter slope of $2$ at lower frequencies to a steeper slope of $3$ at higher frequencies, and the break is connected with the black hole mass. This is similar to the three slope PSD detected in black hole binaries (BHB): $\alpha\sim 0$ for low frequencies ($<0.2$ Hz), $\alpha\sim 1$ for intermediate frequencies ($\sim 0.2-3$ Hz) and $\alpha\sim 2$ at higher frequencies, above $3$ Hz \citep{vaughan2003b}. \\

 \noindent\textbf{(b)Quasi-periodic-Oscillation (QPO):} The origin of QPOs in AGN are highly debated, they are still very rare and they have mostly been discovered in the $2-10\kev$ or harder X-ray bands. For example such QPOs have been found at $2.6\times 10^{-4}$ Hz ($\sim 1$ hour) in RE\,J1034+396 \citep{gierlinski2008,alston2014,alston2016}, at $1.5\times 10^{-4}$ Hz ($\sim 2$ hours) in MS\,2254.9-3712 \citep{alston2015}, at $2.7\times 10^{-4}$ Hz ($\sim 1$ hour) in 1H~0707-495 \citep{pan2016}. A QPO of a period of $\sim 3.8$ hours was detected from an ultra-soft AGN candidate 2XMM J123103.2+110648 \citep{lin2013}. A systematic study of AGN X-ray variability in a sample of 104 sources in search for QPOs detected only two sources with QPOs \citep{gonzalez-martin2012}. Very recently a recurrent QPO has been discovered in the post-changing-look AGN 1ES~1927+654, where the QPO frequency increased from $\sim0.9$ mHz to $\sim 2.3$ mHz over a period of 2 years (Masterson et al. 2025, Nature, in press). \\

\noindent\textbf{(c) X-ray Flares:} X-ray flares with different amplitude at different timescales are common in AGN. 
Typically, flares can exhibit flux increases of $\sim 5-10$ times over time spans ranging from hours to days \citep{gallo2019,lawther2023,wilkins2022,reeves2021, Ding2022}. During X-ray flares, a spectral softening (softer-when-brighter) and a decreasing reflection fraction  have been observed in some AGNs (e.g., Mrk 335 \citealt{gallo2019}, 1H 0707-495 \citealt{wilkins2014}). In Mrk\,335, the decade-long low flux state has been marked by occasional X-ray ``flares," (see Figure \ref{fig:flare} left panel) which have sometimes brightened by a factor of $50$ within a single day \citep{grupe2012,wilkins2015,gallo2019}. In some highly accreting, low SMBH mass AGN, X-ray flares have been linked to the radial expansion of the corona over the accretion disk: the corona appears brighter when it is more extended outward, while it dims when it is compact and located closer to the SMBH (e.g. \citealt{wilkins2014}). Extreme variability and flares in AGN corona suggest that the compact corona must
be a dynamic structure since the time scales for heating and cooling processes for the hot electrons are less than the light crossing time of the corona \citep{fabian2015}, which prevents the system to settle down to an equilibrium. In addition, the plasma properties also change during a flare. For example, \cite{wilkins2022} found that during a flare, the cutoff energy $E_{\rm C}$ of the primary energy continuum  dropped from $140_{-20}^{+100}\kev$ to $45_{-9}^{+40}\kev$. Another example is the Seyfert 1 galaxy I Zwicky 1 that also showed such dramatic changes in the plasma properties, when the corona rapidly cooled from  $E_{\rm C}\sim$200 to $\sim$15 keV within 5 days in January 2020, as caught by \xmm{} and \nustar{}, \citep{Ding2022}. 

 %%%%%%%%%%%%%%%%%%%%%%%%%%%%%%%%%%%

\begin{figure*}
  \centering 

\hbox{
\includegraphics[width=8.5cm,angle=0]{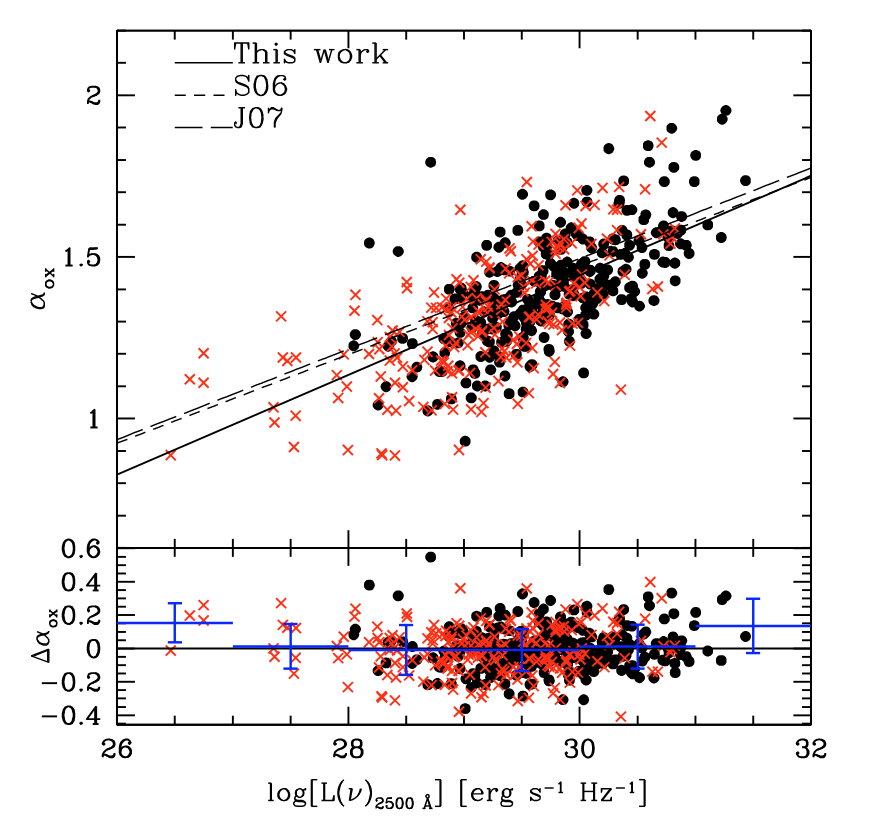}
\includegraphics[width=8.5cm,angle=0]{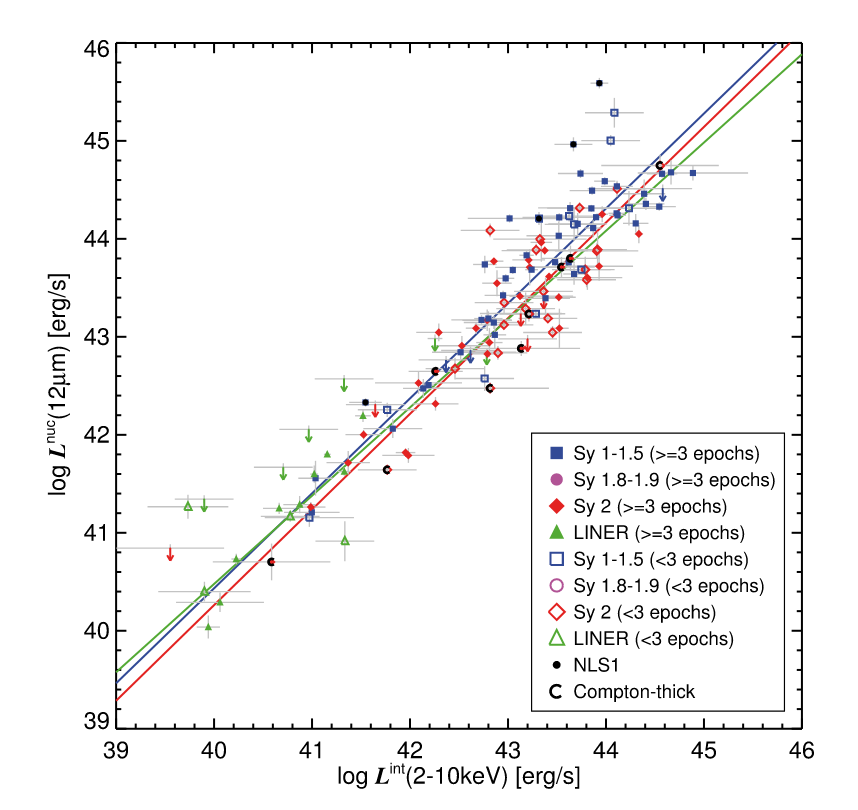}

}\caption{\textbf{Relationship between AGN corona and its surroundings} {\it Left:} The $\alpha_{\rm OX}$ vs $L_{\rm 2500 \rm \AA}$ correlation, a relation connecting the accretion disk emission (UV) and coronal emission \citep{lusso2010}. S06 and J07 refers to \cite{steffen2006} and \cite{just2007} respectively.  {\it Right:} The tight correlation between the spatially resolved nuclear IR emission $L_{12\mu}$ with the coronal X-ray emission \citep{asmus2015}, indicating a close relationship between the corona and the torus. 
} \label{fig:correlations}

\end{figure*}

X-ray flaring events in many astrophysical objects are generally associated with magnetic reconnection \citep[e.g.,][]{petropoulou+16,mehlhaff+20}, a fundamental plasma process where magnetic energy is converted into thermal and nonthermal particle energy \citep[e.g.,][]{lyubarsky05,takahashi+11}. Magnetic reconnection has a short dissipation time, and short flares with durations that generally do not exceed a few times $t\sim R_{\rm G}/c$ should be preferentially associated with magnetic reconnection \citep[e.g.,][]{petropoulou+16,christie+18}. 
Longer flaring episodes could be associated with magnetic flux accumulation in the corona because of accretion, changing the properties of particle acceleration and thus that of the emitted photons \citep[e.g.,][]{liska+20,scepi21,ripperda+22}.

%%%%%%%%%%%%%%%%%%%.  Sec 2.4  %%%%%%%%%%%%%%%%%%%

\subsection{The accretion disk - corona relation}

The accretion disk and the corona are energetically and geometrically related \citep{haardt1993,lusso2010,lusso2016}. A direct piece of evidence of the energy-coupling between the accretion disk and corona is the significant correlation between the quantity $\alpha_{\rm OX}$ and the mono-chromatic UV flux at $2500 \angstrom$ (See Figure \ref{fig:correlations} left panel) \citep{lusso2017_quasars_as_standard_candles}. The parameter $\alpha_{\rm OX}= -0.385\log(F_{2\kev}/F_{2500\rm \angstrom})$ is the ratio between the flux densities at $2\kev$ ($F_{2\kev}$) and $2500 \angstrom$ ($ F_{2500\angstrom}$). Together, the accretion disk and corona form a tightly coupled system.

 There is also a significant correlation between $\alpha_{\rm OX}$ and $\lambda_{\rm Edd}$ \citep{lusso2010}, in which the ratio between X-ray and optical flux decreases with increasing Eddington ratio $\lambdaedd$, implying increased accretion leads to weaker coronal emission. It has also been noted that at sub- and super-Eddington accretion levels, the disc-corona relations are different \citep{huang2020}. For example, the hard X-ray slope $\Gamma$ and the Eddington ratio $\lambdaedd$ show an anti-correlation for sources with lower accretion rate $\lambdaedd <10^{-3}$. See Figure \ref{fig:reverberation} left panel \citep{connolly2016}. On the other-hand a positive correlation is detected for higher Eddington ratio sources, which indicates a softer-when-brighter behavior common in higher accretion rate AGN \citep{mchardy1999,shemmer2006,sobolewska2009}.  Perhaps for weakly accreting AGN ($\lambdaedd <10^{-3}$), the disc-corona system transits to an advection-dominated accretion flow (ADAF), and the X-ray emission may arise from Comptonization process in ADAF \citep{Cao2009}. 

 The reverberation mapping time lags between the optical/UV and the X-rays are an important indication of disk-corona coupling and serves as an important tool to understand the disk-corona geometry \citep[see for e.g.,][and references therein]{peterson_reverb_1993,edelson_reverb_2015,edelson_reverb_2019,cackett_reverb_2021,cackett_reverb_2023,kara_reverb_2023}. The corona is compact $\lesssim 10 R_{\rm G}$ and centrally located relative to the accretion disk, and the UV and optical emission is expected to respond to the incident (and varying) X-ray flux, ``echoing” the X-ray light curve variations after a time
delay corresponding  to the light-travel time across the disk \citep{krolik1991}. For example, for the well studied case of AGN NGC~5548 \citep{fausnaugh_reverberation_NGC5548_2016} significant time delays between the X-rays and the optical-UV band ($1158\angstrom- 9160\angstrom$) have been detected. The trend of lag ($\tau$) with wavelength ($\lambda$) is broadly consistent with the prediction for continuum reprocessing by a thin accretion disk with $\tau \propto \lambda^{4/3}$ (See Figure \ref{fig:reverberation} right panel).

\begin{figure*}
  \centering

\includegraphics[width=18.1cm,height=8cm,angle=0]{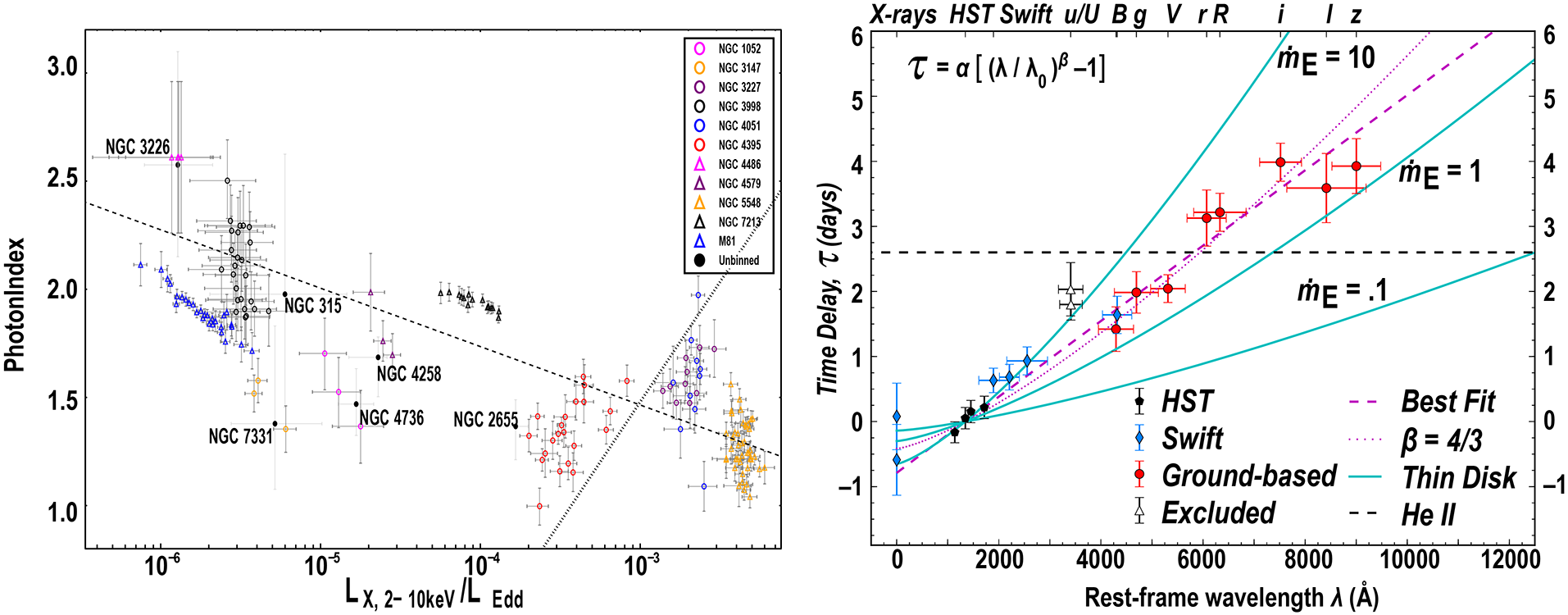}

\caption{\textbf{The relation between the coronal emission and AGN accretion:} {\it Left:} The relation between the photon-indices $\Gamma$ and $L_{2-10\kev}/\ledd$ \citep{connolly2016}. At lower Eddington ratios ($ <10^{-3}$) the quantities show an anti-correlation, while for higher accretion rates the correlation is positive indicating a softer-when-brighter behavior of the corona \citep{Cao2009}. {\it Right:} The X-ray-optical-UV reverberation time lags measured for the source NGC~5548  \citep{fausnaugh_reverberation_NGC5548_2016}. The trend of lag with wavelength is broadly consistent with the prediction for continuum reprocessing by a thin accretion disk with $\tau \propto \lambda^{4/3}$. Here $\dot{m}_{\rm E}=\lbol/\ledd$.} \label{fig:reverberation}
\end{figure*}

 Although the correlation between the UV and X-rays are pretty well constrained in most cases, there are AGN types which show additional complexities. The most interesting among them are changing look AGNs (CL-AGNs), which are sources that undergo a rapid change in flux and spectral state (in optical/X-rays) in a matter of months-years \citep[see][for a review]{ricci_NatAs_review_2023}. Mrk~590 is a long term CL-AGN where the UV and X-rays are well correlated, but UV response to X-ray changes is lagged by $\sim 3$ days indicating a complex reprocessing geometry \citep[a lamp-post geometry would predict a zero time lag, see for e.g.,][]{lawther2023}. In the most enigmatic rapid CL-AGN 1ES~1927+654, the situation is more extreme, with no correlation between the UV and X-ray emission during the violent event \citep{trakhtenbrot2019,laha_1ES_2022, ricci2021}. The right panel of Figure \ref{fig:flare} shows the absolutely uncorrelated behavior of the X-rays and UV for this source. The X-ray coronal emission of 1ES~1927+654 completely vanished a few months after the violent optical outburst, while the UV was still bright and dropping at a rate $\propto t^{-0.91\pm 0.04}$. The corona reappeared after $\sim 4$ months at 10 times the previous luminosity \citep{trakhtenbrot2019,ricci2020,ricci2021,laha_1ES_2022,masterson2022}, finally reaching its normal state in about $\sim 4$ years. In these extreme situations of AGN accretion, the standard disk-corona relations may not hold.

%%%%%%%%%%%%%%%%%%%%%%%%%%%%%%%%%%%%%%%%%%%%%%%%%%

\subsection{Coronae and high-energy neutrinos}
 
 Systematic searches for neutrino excess above atmospheric and cosmic backgrounds with the {\it IceCube} detector have detected $79_{-20}^{+22}$ neutrinos at TeV energies from the nearby AGN NGC~1068 with a significance of $\sim 4.2\sigma$ \citep{IceCube-NGC1068}. Notably, the isotropic neutrino luminosity ($L_{\nu}=2.9_{-1.1}^{+1.1}\times 10^{42}\lunit$) in the 1.5–15 TeV range exceeds both the gamma-ray luminosity ($L=1.6\times 10^{41}\lunit$) in the 100 MeV–100 GeV range and the upper limits on gamma-ray luminosity above 200 GeV \citep{aartsen+20,IceCube-NGC1068}. This suggests the AGN's central engine (the X-ray corona), which is opaque to gamma-rays, significantly contributes to neutrino production \citep[e.g.,][]{fang+23,murase+20,inoue+20,eichmann+22,mbarek+24,fiorillo+24,padovani+24}. Other AGNs also show hints of neutrino emission, with future deeper follow-ups expected to enhance detection significance \citep{neronov+23,murase+24}. Beyond AGNs, X-ray binary coronae have been proposed as potential sources of Galactic neutrinos detected by IceCube \citep{fang+24}.  

The highly magnetized black hole coronae \citep[e.g.,][]{beloborodov17,hooper+23,groselj+24,nattila24,mbarek+24} support two primary mechanisms for accelerating protons responsible for coronal neutrino production. First, magnetic reconnection in large current sheets near the black hole, with strong guide fields, can accelerate particles to extreme energies \citep{fiorillo+24}. Second, magnetized turbulence allows particles to be initially energized by reconnection and subsequently re-accelerated within the turbulent corona \citep{mbarek+24}. These high-energy protons interact with the corona’s dense photon fields, producing the observed neutrino signal. While purely leptonic models have also been suggested \citep{hooper+23}, it remains unclear how electrons could achieve the TeV-scale energies required for such scenarios.

%%%%%%%%%%%%%%%%%%%. %%%%%%%%%%%%%%%%%%%. %%%%%%%%%%%%%%%%%%%. %%%%%%%%%%%%%%%%%%%. 

 \begin{table*}[]
    \centering
    \caption{The empirical relations involving AGN coronal emission.\label{tab:emperical}}
    \begin{tabular}{lccccc}
    \hline \hline
             
        Relationship between \hspace{1cm} & Equation   \\\hline
        
        (1a)X-ray and  UV    & $\alpha_{\rm OX}=(0.154\pm-0.010)\log L_{2500 \rm \AA} -(3.176\pm0.223)$ \\
        
        (1b) X-ray and  UV       & $L_X \propto L_{\rm UV}^{0.7-0.8} $\\
        
        (2a) $\Gamma$ vs $\lambdaedd{}$ (for $\lambdaedd>0.01$)& $\Gamma=(0.41\pm0.09)\log \lambdaedd + (2.17\pm0.07)$                                               \\

        (2b) $\Gamma$ vs $\lambdaedd{}$ (for $\lambdaedd<0.01$ )& $\Gamma=(-0.09\pm0.03)\log \lambdaedd + (1.55\pm0.07)$                                               \\
        
        (3) $2-10\kev$ and  Infrared & $\log \frac{L^{\rm int}_{2-10\kev}}{10^{43}\lunit}= (-0.32\pm 0.03) + (0.95\pm 0.03) \log \frac{L_{12\mu}^{\rm nuc}}{10^{43}}\lunit{}$    \\
        
        (4) $L_{2-10\kev}$ vs OIII &  $\log L_{2-10\kev}= 0.95 \log L_{OIII} + 3.89$\\
        
        %(5)LX-LUV-LRadio & \\
        
        (5) Fundamental plane of black hole & $\log L_{\rm R}=(0.60\pm0.11)\log L_{\rm X} + (0.78\pm 0.11)\log M + (7.33\pm 4.05)$\\ 
        
        (6a) Gudel Benz relation& $L_{R, \, 5GHz}\sim 10^{-5.5} L_{2-10\kev}$ \\
        
        (6b) 100 GHz vs 2-10 $\kev$   & $\log L_{100 \rm GHz}$  = $(-13.9 \pm 0.8) +(1.22\pm 0.02) \log L_{2-10\kev}$\\
        
        (7) Iwasawa-Taniguchi effect  & $\log({\rm EWFeK}\alpha)\propto (-0.17\pm0.03) \log(L_{2-10\kev})$\\

        %(7) $L_{2\kev}$ vs $L_{2500 \rm \AA}$ & $L_{2\kev} \propto L_{2500 \rm \AA}^{0.6-0.8}$& \cite{lusso2016}\\
        
        \hline
    \end{tabular}
    {References: (1a):\cite{lusso2010,lusso2016},
    (1b): \citep{just2007,strateva2005}, 
    (2a): \cite{risaliti2009,kelly2008,shemmer2008},
    (2b):\cite{gu2009},
    (3): \citep{gandhi2009,asmus2015}
    (4) : \cite{saade2022,malkan2017}
    %(5): \cite{zhu2020}
    (5): \cite{merloni2003}
    (6a):\cite{laor2008}
    (6b):\cite{ricci_radio_correl_2023}
    (7): \cite{iwasawa1993}}

    { (2a) This positive correlation exists for accretion rates $\lambdaedd >0.01$ (2b)For low-luminosity AGN, hence low accretion states, there exists an anti-correlation.}
    
    {Note that here X-ray, $L_{\rm X-ray}$ and $L_{2-10\kev}$ has been interchangeably used and can be interpreted as similar quantity.}

\end{table*}

%%%%%%%%%%%%%%%%%%%.  Sec 2.12 %%%%%%%%%%%%%%%%%%%

\subsection{The empirical relations involving corona}

Here we list the most important empirical relations involving emission from the X-ray corona and that produced by other AGN components (see Table \ref{tab:emperical} for a list).\\

\noindent\textbf{(1) X-ray and UV:} As noted earlier, the disk and the corona emission are very tightly related, showing a strong correlation between $\alpha_{\rm OX}$ vs $L_{\rm 2500\angstrom}$ \citep{lusso2010,bisogni2021}. \\

\noindent\textbf{(2) $\Gamma-\lambdaedd$:} A correlation exists between $\Gamma$ and $\lambdaedd$ ($\Gamma\sim 0.3 \times \log \lambdaedd +2$) such that sources with $\lambdaedd>0.3$ have a very steep slope of $\Gamma>2$ \citep{shemmer2008,risaliti2009,brightman2013}. The correlation can be explained as increased UV emission from the accretion disk due to high accretion rate can lead to radiative cooling of the X-ray corona and hence lowering of the electron temperature (that is a steeper X-ray spectrum). The left panel of Fig \ref{fig:reverberation} shows this behavior between $\Gamma$ vs $L_{2-10\kev}/\ledd$, where we find that the higher accreting sources show a positive correlation between the two quantities, while for very low accreting sources, there is an anti-correlation \citep{fausnaugh_reverberation_NGC5548_2016,gu2009}. By simulating AGN populations with an X-ray spectral Comptonization model, \cite{ricci2018} showed that Comptonizing plasma with temperatures and compactness lying along the pair line can straightforwardly explain the positive correlation between $\Gamma$ and the Eddington ratio. 

 A few studies involving high $\lambdaedd$ sources (with $\lbol=10^{46}\lunit$) did not detect any correlation between $\Gamma$ and $\lambdaedd$  \citep{laurenti2022,liu2021}. We note here that these highly accreting sources have higher radiation pressure from the accretion disk which could affect the structure and efficiency of the accretion disc-corona system, and hence one would expect different behavior between UV and X-rays. Similar lack of $\Gamma-\lambdaedd$ has been found in a hard X-ray study of a sample of nearby AGN \citep{kamraj2022}. \\

%%%%%%%%%%%%%%%%%%%%%%%%%%%%%%%%%%%%%%%%%%%%%

\noindent\textbf{(3) $L_{12\mu}$  vs $L_{\rm 2-10\kev}$:} The spatially resolved core IR luminosity at 12$\mu$ is correlated very strongly with the $2-10\kev$ X-ray emission \citep{gandhi2009,asmus2015}. See Figure \ref{fig:correlations} right panel. \\

\noindent\textbf{(4) $L_{\rm 2-10\kev}$ vs OIII:} The hard X-rays are correlated with the optical emission line intensity \citep{bassani1999,malkan2017,saade2022}. \\

\noindent\textbf{(5) Fundamental plane of BH:}  The X-ray luminosity ($L_{2-10\kev}$), the radio  luminosity ($L_{5\rm GHz}$), and central black hole mass ($\mbh$) in accreting systems have long been suggested to be interrelated \citep{tananbaum1983}, and this connection has since been further established \citep{worrall1987,merloni2003,miller2011,zhu2020}.  The relation between these three quantities potentially serves as an indicator of similar physics in action across different mass scales of accreting systems. However, recent studies with very high spatial resolution in radio band found that the the core radio luminosity ($L_{5\rm GHz}$) is consistently lower than that predicted by the fundamental plane relation, and the relation only holds true if we consider the extended radio emission from the host galaxy \citep{fischer2021}. \\

\noindent\textbf{(6) The mm and X-ray relation:} As mentioned earlier, AGN show  a strong correlation between their X-ray and radio luminosity at 5-200\,GHz with $L_{\rm R}/L_{\rm X-ray}\sim 10^{-5.5}$ \citep{ricci_radio_correl_2023,behar2015,behar2018,kawamuro2022}. As noted in the introduction, at lower radio frequencies ($<45$ GHz) synchrotron self absorption prevents a direct view of the coronal radio emission. \\

\noindent\textbf{(7) $L_{\rm 2-10\kev}$ vs H$\beta$-FWHM and H$\alpha$ luminosity:}  It has been observed that the X-ray  luminosity correlates well with the broad H$\beta$ full-width-at-half-maximum (FWHM) and the H$\alpha$ luminosity. \citep{laor1997,brandt1997,shemmer2006}. \\

\noindent\textbf{(8) The Iwasawa-Taniguchi effect:} Also known as the X-ray Baldwin effect, is the anti-correlation between the equivalent width of the Fe~K$\alpha$ emission and the $2-10\kev$ flux \citep{iwasawa1993,page2004}.

%%%%%%%%%%%%%%%%%%%.  Sec 4  %%%%%%%%%%%%%%%%%%%

\section{List of open questions}

Although we have discussed numerous observational discoveries regarding AGN coronal emission, several fundamental questions continue to elude us. We list some of the outstanding questions below, that, if answered, will improve our understanding of, not only the corona, but also how AGN operate.

\begin{itemize}

   \item {\it Since corona is found ubiquitously in AGN, is there something fundamental about the accretion process that produces it?}
     The corona is a unique physical entity found in most accreting systems including black hole binaries (BHBs) and AGN. Studies have found similarities in coronal behavior of AGN and BHBs, lying at the two ends of black hole mass scales, suggesting that the AGN accretion-disk and corona are just a scaled-up version of those found in BHBs \citep{mchardy2006}, with the underlying physics being the same. Possibly the magnetic fields that thread the accretion disk creates and sustains the corona in these accreting systems, whose physics remains similar across a large range of black hole masses ($\sim 10- 10^9 \msol$).

    \item {\it What is the geometry of the corona? }
The recent X-ray polarimetric results with {\it IXPE} point towards a more extended geometry of the corona, situated along the accretion disk plane. However, deeper polarimetric studies of larger samples of AGN in different flux states are required to understand how the geometry varies depending on accretion and X-ray luminosity states.  This can only be done with the next generation of X-ray polarimeters, as {\it IXPE} is sensitivity-limited. A systematic spectral and timing studies of the AGNs in rapidly changing X-ray flux states can also reveal the geometry.

    \item {\it What are the main energy pumping and dissipation mechanisms in the corona? Is the corona in thermal and radiative equilibrium?} Although random magnetic reconnection events can play an important role in pumping energy into the corona \citep{galeev1979, dimatteo1998,merloni2001_accdisk,sironi2020,sridhar2021}, we need a deeper understanding about heating and cooling processes in such a compact region, which shows constant stochastic fluctuations, and sometimes flares. Simulations coupled with observational inputs on simultaneous radiative and thermal equilibrium can shed light on this topic in the future.

    \item  {\it What determines the fraction of non-thermal electrons in the X-ray emitting plasma?}
 As mentioned in the introduction, the existence of non-thermal particles in the corona would result in a distribution of photons that extends into the the MeV band. This small number of high energy particles could be highly effective in seeding pair production and can share the total available energy, thus reducing the mean energy per particle and therefore decreasing the temperature of the thermal population. Thus the non-thermal fraction of particles in the corona plays an important role in balancing the temperature of the plasma. We do not understand the origin and the exact fraction of the non-thermal electrons in coronal plasma. Future simulations on magnetic reconnection events could help us understand this.

    \item {\it What are coronal flares?} It is not clear to us if there are particular flux/spectral states that favor the occurrence of coronal flares (rise in flux by a factor of $\gtrsim 5$ times in a few days/months). It is also still unclear if the X-ray flares are the main energy dissipation mechanisms in the corona?

\end{itemize}

\section{Future perspectives}

\subsection{Need for future missions}
 
Future X-ray studies on AGN corona depends on how well we can extend our spectroscopic capabilities in the hard X-rays, preferably up to $\sim 500\kev$. Currently \nustar{} has a bandpass up to $79\kev$ and constraining cut-off energy $>100\kev$ becomes highly uncertain and model dependent. For example, the $E_C$ estimated for NGC~5506 using the same \nustar{} observation by two different models found $E_{\rm C}=720\pm130\kev$ \citep{matt2015} and $110\pm10\kev$ \cite{balokovic2020}. 

 To estimate the differences in the emissivity profiles  (the illumination pattern of the accretion disk due to the reflection of X-rays from the corona, convolved with general relativistic effects), and hence the coronal shape and size, we need high quality X-ray observations, both in terms of collecting area and spectral resolution.
For example, missions like {\it Athena} with its large collecting area \citep{athena_nandra_2013} and the recently launched {\it XRISM} with its high spectral resolution \citep{xrism_tashiro_2020} will provide the ability to distinguish between the different coronal geometries. 

The exciting field of X-ray polarimetry has just taken off with the launch of IXPE. However, a small effective-area mission such as this one needs much longer integration time to constrain the polarisation degree for even a very bright AGN ($\sim 500\ks$ needed for MCG-5-23-16 to obtain an upper limit on polarization degree \cite{marinucci_IXPE2022}). Future X-ray polarimeters should have very large effective area not only to constrain polarization parameters at a fraction of the exposure required by IXPE, but also carry out time dependent polarimetric analysis of AGN.

 High-energy neutrinos are also expected from AGN corona, and is currently opening up a huge multi-messenger avenue for AGN-coronal studies \citep{neutrino_corona_2021}. In the future, deeper and more sensitive studies by {\it IceCube} and other detectors will help us in understanding the relation between neutrino emission and the physical processes in an AGN corona.

\subsection{Need for simulations}

Coronal heating and cooling problems are among the most significant unresolved issues in astrophysics. `Fluid' (MHD) models, by their very nature, are unable to explore the physics of non-thermal particle acceleration within the dissipation regions, where the energy from the magnetic field is transferred to particles. The corona is expected to have a good fraction (up to $30\%$) of non-thermal particles \citep{fabian2017}. In MHD simulations, the energy either stays in the system as thermal energy in the particles or is removed according to some ad-hoc prescription. This is because, most of the non-thermal acceleration occurs in `collisionless' plasmas, where Coulomb collisions, typically an efficient means of thermalization, are explicitly neglected. Therefore, the properties of the population of non-thermal particles responsible for the emission cannot be properly captured in fluid models. 

On the other hand, particle-in-cell (PIC) simulations capture the microscopic dynamics of individual particles, rather than assuming a smooth distribution of particle energies, and thus capture accurately the non-thermal processes in dissipation regions, and the nonlinear interplay between charged particles and electromagnetic fields \citep[e.g.,][]{chernoglazov+23,groselj+24,mbarek+24,nattila24}. Moreover, PIC simulations may include consistently evolving particles and their radiative cooling effects, in the presence of photons, pair creation and annihilation processes \citep{groselj+24}. These features make PIC an ideal tool to study the coronal heating problem. However, PIC simulations are usually employed to study local dissipation processes on microscopic scales---scaled down from actual astrophysical scales. Therefore, simulation setups might seem ideal and somewhat disconnected from the `global properties' of the corona. More work is still required to more robustly extrapolate the results of PIC simulations to large scales \citep[underway efforts include, e.g.,][]{zou+24,sridhar+24}.

%%%%%%%%%%%%%%%%&%%%%%%%%%%%%%%%%%%%%%%%%%%%%%%%%%%%%%%%%%%%%%%%%%%%
                
\acknowledgements
The material is based upon work supported by NASA under award number 80GSFC21M0002. SL acknowledges insightful discussions with Christopher Reynolds, Mitchell Begelman,  Navin Sridhar and Dev Sadaula. SL thanks NASA graphics designer Jay Friedlander for his help on the cartoon in Figure 1 and other figures. CR acknowledges support from Fondecyt Regular grant 1230345, ANID BASAL project FB210003 and the China-Chile joint research fund. E.B. acknowledges support by a Center of Excellence of the Israel Science Foundation (grant no. 2752/19). SL and EB acknowledge support from NSF-BSF grant numbers: NSF-2407801, BSF-2023752.

\bibliographystyle{aasjournal}
\bibliography{mybib}
\end{document}